%% file: Main.tex
\documentclass[journal,onecolumn]{IEEEtran}

\input{Packages.tex}
\include{mymacro}

\def\BibTeX{{\rm B\kern-.05em{\sc i\kern-.025em b}\kern-.08em
    T\kern-.1667em\lower.7ex\hbox{E}\kern-.125emX}}
\begin{document}

\title{Preliminaries paper: Byzantine Tolerant Strong Auditable Atomic Register}

\author{\IEEEauthorblockN{Antonella Del Pozzo}
\IEEEauthorblockA{
Université Paris-Saclay, CEA, List,\\ F-91120, Palaiseau, France \\ antonella.delpozzo@cea.fr \\}
\and
\IEEEauthorblockN{Antoine Lavandier}
\IEEEauthorblockA{
Université Paris-Saclay, CEA, List,\\
F-91120, Palaiseau, France \\
antoine.lavandier@cea.fr \\}
\and
\IEEEauthorblockN{Alexandre Rapetti}
\IEEEauthorblockA{
Aix-Marseille University, \\ Université Paris-Saclay, CEA, List,\\
F-91120, Palaiseau, France \\
alexandre.rapetti@cea.fr}
}

\maketitle

\begin{abstract}
\input{abstract}

\end{abstract}


\section{Introduction}
\input{introduction}
\label{sec:intro}


\section{System Model}
\label{sec:systemModel}
\input{systemModel}

\section{Single-Writer/Multi-Reader Atomic Auditable Register}
\label{sec:interface}
\input{interface}

\section{Impossibility results}
\label{sec:impossibleResult}
\input{impossibilty}


\section{Solution specification}
\label{sec:asynchAlgorithm}
\input{asyncUnboundedSolution}

\section{Valid reads using multiple writers}
\label{sec:validRead}
\input{writer_quorum}


\bibliographystyle{IEEEtran}
\bibliography{IEEEabrv,references}















\end{document}

%% file: Packages.tex
\usepackage{cite}
\usepackage{amsmath,amssymb,amsfonts}
\usepackage{graphicx}
\usepackage{textcomp}
\usepackage{xcolor}
\usepackage{tikz}
\usepackage[a4paper,top=2cm,bottom=2.25cm,left=1.75cm,right=1.75cm]{geometry} 

\usepackage[colorlinks=true, allcolors=blue]{hyperref}
\usepackage[ruled]{algorithm}
\usepackage[noend]{algpseudocode}
\usepackage{tabularx}
\usepackage[shortlabels]{enumitem}

%% file: mymacro.tex
\definecolor{g0}{HTML}{339900}
\definecolor{o0}{HTML}{FF6600}
\definecolor{b0}{HTML}{0033CC}

\newcommand{\reg}[1]{R_{#1}}

\newcommand{\eventSeq}{H}
\newcommand{\signature}{\Sigma}

\newcommand{\writets}{WRITE} 
\newcommand{\writetsack}{WRITE\_ACK} 

\newcommand{\getts}{TS\_REQ}
\newcommand{\TS}{TS\_RESP}
\newcommand{\getval}{VAL\_REQ}
\newcommand{\VALS}{VAL\_REP}

\newcommand{\AUDIT}{AUDIT}
\newcommand{\AuditResp}{AUDIT\_RESP}
\newcommand{\AuditReq}{AUDIT\_REQ}

\newcommand{\GetBlocks}{BLOCK\_REQ}
\newcommand{\SendBlock}{BLOCK\_RESP}
\newcommand{\GetValUnbound}{VAL\_REQ}
\newcommand{\SendValUnbound}{VAL\_RESP}

\newcommand{\invokeAudit}[2]{auditReq_{#2}(#1)}
\newcommand{\returnAudit}[3]{auditRep_{#2}(#1,#3)}

\newcommand{\RVal}{val}

\newcommand{\Rts}{reg\_ts}

\newcommand{\decrypt}{decrypt}

\newcommand{\READERS}{Reads\_temp}
\newcommand{\ReadsServer}{Pending\_reads}

\newcommand{\writerAcks}{Acks}

\newcommand{\collectedValues}{Collected\_blocks}

\newcommand{\collectedTs}{Collected\_ts}

\newcommand{\nSeq}{n\_seq}

\newcommand{\generatesBlock}{GenerateBlocks}

\newcommand{\GetValue}{GetValue}

\newcommand{\notOld}{notOld}
\newcommand{\ValidBlock}{validBlocks}

\newcommand{\smallestTimestamp}{min\_ts}

\newcommand{\collectedLog}{Collected\_log}
\newcommand{\Logi}{Log}

\newcommand{\pss}{{s}}
\newcommand{\ps}[1]{{s_#1}}
\newcommand{\pw}{p_{w}}
\newcommand{\pri}[1]{p_{r_#1}}
\newcommand{\pr}{p_{r}}
\newcommand{\pa}{p_{a}}
\newcommand{\pR}{p_{R}} 
\newcommand{\FaultySet}{\mathcal{B}}

\newcommand{\tend}[1]{t_{end}(#1)}
\newcommand{\tbegin}[1]{t_{start}(#1)}

\newcommand{\op}[1]{Op_{#1}}
\newcommand{\opts}[1]{\op{#1}.ts}

\newcommand{\spaceLemmaProof}{\vspace{0.5em}}

\newtheorem{theorem}{Theorem}
\newtheorem{lemma}{Lemma}

\newtheorem{observation}{Observation}

\newcommand{\toto}{xxx}
\newenvironment{proofT}{\noindent{\bf
Proof }} {\hspace*{\fill}$\Box$\par\vspace{3mm}}
\newenvironment{proofL}{\noindent{\bf
Proof }} {\hspace*{\fill}$\Box$\par\vspace{3mm}}

\algblock{Phase}{EndPhase}
\algblockdefx[Phase]{START}{}%
    [1][If]{\textbf{#1}}%
\algnotext[Phase]{EndPhase}

%% file: abstract.tex
An auditable register extends the classical register with an audit operation that returns information on the read operations performed on the register. In this paper, we study Byzantine resilient auditable register implementations in an asynchronous message-passing system. Existing solutions implement the auditable register on top of at least 4f+1 servers, where at most $f$ can be Byzantine. We show that 4f+1 servers are necessary to implement auditability without communication between servers, or implement does not implement strong auditability when relaxing the constraint on the servers' communication, letting them interact with each other. In this setting, it exists a solution using 3f+1 servers to implement a simple auditable atomic register. In this work, we implement strong auditable register using 3f+1 servers with server to server communication, this result reinforced that with communication between servers, auditability (event strong auditability) does not come with an additional cost in terms of the number of servers.

%% file: introduction.tex
Outsourcing \emph{storage} capabilities to third-party distributed storage are commons practices for both private and professional users. This helps to circumvent local space limitations, dependability, and accessibility problems. However, this opens to problems such as users that have to trust the distributed storage provider on data integrity, retrievability, and privacy. As emphasized by the relentless attacks on servers storing data \cite{DB18} and by the recent worldwide advent of data protection regulations\cite{gdpr,ccpa,pipl}.

In this work, we address the problem of bringing \textit{auditability} in a distributed storage system, i.e., the capability of detecting who has read the stored data. 
We consider a set of servers implementing the distributed storage and a set of clients (users) accessing it through read and write operations. Auditability implies the ability to report all the
read operations performed by clients. Nevertheless, without reporting
any client that did not read. Let us note that once a reader
accesses a value, it can disclose it directly without
being auditable. For that reason, auditability does
not encompass this kind of behavior.

\subsection{Related work.} 

Most of the results of this work have already been published in~\cite{del2022byzantine}. In this article, we propose a new algorithm that implements a Strong Auditable Atomic Register (with completeness and strong accuracy) using 3f+1 servers, while in~\cite{del2022byzantine}, the solution provides only an auditable atomic register (with completeness and accuracy).

\subsection{Our Contribution}
Our contributions are the following:
\begin{itemize}
    
  \item A new algorithm implementing a strong auditable atomic register with $3f+1$ servers.

  \item An experimentation in rust using Zenoh~\cite{desbiens2022zenoh} of this algorithm.

\end{itemize}

{\bf Paper organization.} The paper is organized as follows. Section \ref{sec:systemModel} defines the system model. Section \ref{sec:interface} formalizes the auditable register abstraction and properties. Section \ref{sec:impossibleResult} gives a lower bound on the number of servers to implement an auditable register. Section \ref{sec:asynchAlgorithm} presents an optimal resilient algorithm implementing the Auditable Atomic Register and gives the proof of its correctness. Section~\ref{sec:validRead} presents an approach to consider multiples writer, tolerating byzantine failure, in a special context where all correct writers aims to write the same value.



%% file: systemModel.tex
We consider an asynchronous message-passing distributed system composed of a finite set of sequential processes. Each process has a unique ID and is equipped with a cryptographic primitive $\signature$ to sign the messages it sends. We assume that signatures are not forgeable. 
A process can be either a \textit{client} or a \textit{server}. We consider an arbitrary number of clients and $n$ servers that implement a distributed register. The writer is a special client that owns the register and is the only one allowed to write on it. The other clients are the readers, who can read the register's content. In the following, we denote the readers as $\pri{1}, \pri{2}, \dots$, the writer (and auditor) of the register as $\pw$ and the servers as $\ps{1},..., \ps{n}$.

\subsection{Failure model}
We consider that all correct processes follow the same protocol $A$. A process that executes any algorithm $A'\neq A$ is considered Byzantine. The writer can only fail by crashing. At most $f$ servers and any number of readers can be \textit{Byzantine}.  However, we consider that any Byzantine faulty reader does not cooperate with other faulty processes.


\subsection{Communication primitives}
The writer broadcasts messages to servers using a reliable broadcast primitive \cite{bracha85}. The broadcast is done by invoking the \texttt{broadcast} primitive and the delivery of a broadcast message is notified by the event \texttt{deliver}. This primitive provides the following guarantees: \textbf{Validity:}\textit{ If a correct process \texttt{deliver} a message $m$ from a correct process $p_i$, then $p_i$ \texttt{broadcast} $m$;} \textbf{Integrity:} \textit{No correct process \texttt{deliver} a message more than once;} 
\textbf{No-duplicity:} \textit{No two correct processes \texttt{deliver} distinct messages from $p_i$;} 
\textbf{Termination-1:} \textit{If the sender $p_i$ is correct, all the correct processes eventually \texttt{deliver} its message;}
\textbf{Termination-2:} \textit{If a correct process \texttt{deliver} a message from $p_i$ (possibly faulty) then all the correct processes eventually brb-deliver a message from $p_i$.}

Processes can communicate with each other using a perfect point to point communication abstraction. Processes send messages by invoking the \texttt{send} primitive and the reception of a message is notified by the event \texttt{receive}. This abstraction provides the following guarantees:  \textbf{Reliable delivery:} \textit{If a correct process $p_i$ sends a message $m$ to a correct process $p_j$, then $p_j$ eventually delivers $m$;}  \textbf{Integrity:} \textit{No correct process \texttt{receive} a message more than once;}  \textbf{No creation:} \textit{If some process $p_i$ receives a message $m$ with sender $p_j$, then $m$ was previously sent to $p_i$ by process $p_j$.}

%% file: interface.tex
In this work we define a Single-Writer/Multi-Reader auditable atomic register, that can be transposed to the multi-writer multi-reader case\cite{CGR11}. In the following we first recall the atomic register specification and later we extend it with auditability. 

A register $\reg{}$ is a shared object that provides the processes with two operations, $\reg{}.write(v)$ and $\reg{}.read()$. The first allows to assign a value $v$ to the register, while the second allows the invoking process to obtain the value of the register $\reg{}$.  Being the register a shared object, it can be concurrently accessed by processes and each operation is modeled by two events, an invocation and a response event.

We consider a single-writer/multi-reader atomic register, that can be written by a predetermined process, the \textit{writer}, and read by any processes, the \textit{readers}. Intuitively, atomicity provides the illusion that all the read and write operations appear as if they have been executed sequentially.

The interaction between the processes and the register is modeled by a sequence of invocation and reply events, called a history $H$. Without loss of generality, we assume that no two events occur at the same time. An operation is said to be complete in a history $H$, if $H$ contains both the invocation and the matching response for this operation. If the matching response is missing, the operation is said to be \textit{pending}.

A history is sequential if each operation invocation is followed by the matching response. For a given history $H$ we denote ${\sf complete}(H)$ the history obtained starting from $H$ by appending zero or more responses to pending invocations and discarding the remaining pending invocations. 

A history $H$ is atomic if there is a sequential history $\pi$ that contains all operations in ${\sf complete}(H)$ such that:
\begin{enumerate}
    \setlength\itemsep{-0.1em}
    \item 
     \textit{Each read $\in \pi$ returns the value of the most recent preceding write, if there is one, and otherwise returns the initial value.}
    \item \textit{If the response of an operation $\op{1}$ occurs in ${\sf complete}(H)$, before the invocation of operation $\op{2}$, then $\op{1}$ appears before $\op{2}$ in $\pi$}
\end{enumerate}
Moreover, a history $H$ is wait-free if every operation invoked by a correct process has a matching response.
All the histories generated on Atomic Register are \textbf{atomic} and \textbf{wait-free}.\\

We now define the \textit{auditable atomic register} extending the atomic register with the {\sf audit}$()$ operation and defining its semantics. Let us recall that only the writer can perform that operation. The  {\sf audit}$()$ operation invocation is $\invokeAudit{\reg{}}{}$, and its response is $\returnAudit{\reg{}}{}{Eaudit}$, with $Eaudit$ the list of couples process-value $(p,v)$ reported by the audit operation.

As shown in \cite{CB21}, it is not possible to implement an audit operation in the presence of Byzantine servers, if a single server stores a full copy of the value. Informally, a Byzantine reader could contact only the Byzantine servers, getting the value without leaving any trace to be detected. 

A possible solution to this issue, as presented in \cite{K93}, is to combine secret sharing for secrecy \cite{S79} and information dispersal for space efficiency \cite{R89}.
When writing a value $v$, the writer does not send the whole value to each server, but generates a random key $K$ and encrypts $v$ with it. Then, for space efficiency, the writer uses information dispersal techniques to convert the encrypted value in $n$ parts, ${v_1},{v_2},\ldots,{v_n}$, of size $\frac{|v|}{\tau}$ ($\tau$ is the number of parts needed to reconstruct the value). Finally, the writer uses secret sharing techniques to convert the key $K$ in $n$ shares, $sh_1, sh_2,\ldots sh_n$, such that the share $sh_i$ is encrypted with the public key of the server $s_i$. At this point, the writer can send to the servers $(v_1,sh_1),\dots, (v_n,sh_n)$. Each server stores only its block and decrypted share. 
The secret sharing scheme assures that (1) any $\tau$ shares are enough for a reader to reconstruct the key $K$, and so the value, (2) that less than $\tau$ shares give no information on the secret. 
Those techniques use fingerprints to tolerate alterations by faulty processes and allow reading processes to know when they collect $\tau$ valid blocks to reconstruct the value.
\\
For sake of simplicity in the presentation of our solution, we avoid the details of the secret sharing scheme implementation.
We consider that for any value $v$, the writer constructs a set of blocks $\{b_i=(v_i,sh_i)\}_{i\in[1,n]}$, such that a block $b_i$ can only be decrypted by a server $s_i$. Any $\tau$ blocks are necessary and sufficient to reconstruct and read the value $v$.\label{issue:systemModel:infodispersal}

We use the notion of effectively read, introduced in \cite{CB20}. This notion captures the capability of a process to collect those $\tau$ blocks to reconstruct a value regardless it returns it or not i.e., the corresponding response event may not appear in the history. 

We consider the execution $E$, obtained by adding to the history $H$ the communication events: {\sf send}, {\sf receive}, {\sf broadcast} and {\sf deliver}.

\textbf{Effectively read:} 
\textit{ A value $v \in \mathcal{V}$ is effectively read by a reader $\pr$ in a given execution $E$ if and only if $\exists$ the invocation of a write(v) operation $\in E$ and receive($b_{v_j}$) events for $\tau$ different blocks.
}

We can now define the \textit{auditability} property as the conjunction of the completeness property and the accuracy property.

\begin{itemize} 
\item \textbf{Completeness \cite{CB20} :}
\textit{For every value $v$ written into the register, every process $p$ that has \textsf{effectively read} $v$, before the invocation of an audit operation $op$ in $\eventSeq$, $p$ is reported by $op$ to have read $v$.} 
\label{property:completeness}

\item \textbf{Strong Accuracy:} 
\textit{A correct process $p$, that never \textsf{effectively read} the value $v$, will not be reported by any audit operation to have read $v$.
}
\label{property:StrongAccuracy}
\end{itemize}

The completeness property assures that if a reader $p$ succeeds in obtaining a value $v$ before the invocation of the audit operation, then the $Eaudit$ list will contain the couple $(p, v)$.
The strong accuracy property assures that if a correct reader $p$ never effectively read $v$, then the $Eaudit$ list will never contain the couple $(p, v)$.

In this paper, we propose an optimal resilient solution of the Single-Writer Multi-Reader Strong Auditable Atomic Register. 

In the following, we denote the {\sf read}$()$ (resp. the {\sf write}$()$ and {\sf audit}$()$) operation  to the register as $\op{r_i}$ (resp. $\op{w_i}$ and $\op{a_i}$).


%% file: impossibilty.tex
In this section, we first recall an impossibility result from \cite{CB20} that provides a necessary condition on the number of blocks $\tau$ to have auditability, that we extend in our system model. 
Finally, we show that without communication between servers, it is impossible to implement an auditable register with less than $4f+1$ processes.
Hereafter, the impossibility result presented in \cite{CB20} with the complete proof in our system model.

\begin{theorem}
\label{theorem:impossible:bessaniCompleteness}
Let $\tau$ be the number of blocks necessary to recover a value written into the register $\reg{}$. In presence of $f$ Byzantine servers, it is impossible to provide completeness if $\tau<2f+1$.
\end{theorem}\spaceLemmaProof

\begin{proofT}
 Let $c$ be a Byzantine client. To read a value from the register, $c$ needs to collect $\tau$ blocks. 
In the following, we show that, if $\tau<2f+1$, a client $c$ can read a value $v$ and $\langle c,v\rangle$ is not returned by any audit operation. 
Consider the execution where
$c$, during the execution of read operation $\op{}$, obtains the $\tau$ blocks from $f$ Byzantine servers denoted $S_1$, $\tau-2f$ correct servers denoted $S_2$ and $f$ other correct servers denoted $S_3$. The  remaining $n-\tau$ correct servers, denoted $S_4$, have no information about $\op{}$.

An audit operation that starts after $\op{}$ returns cannot wait for more than $n-f$ responses from the servers. It is possible that those responses are the ones from $S_1 \cup S_2 \cup S_4$. $S_1$ being the Byzantine, do not report $\op{}$. Processes in $S_4$ have no information about the read operation of $c$. Then there are only the $\tau-2f$ servers of $S_2$ that report process $c$. 
Since $\tau<2f+1$, there is no server that can report process $c$ to have read $v$.

\renewcommand{\toto}{theorem:impossible:bessaniCompleteness}
\end{proofT}

Intuitively, the value of $\tau$ has to be sufficiently big to (i) impede $f$ Byzantine servers from collaborating and reconstructing the value and (ii) to force a reader when reading to contact sufficiently many correct servers to be auditable for that operation. Thus, the number of blocks $\tau$ also corresponds to the number of servers that must be contacted to read.
Without loss of generality, in the following we consider that each server stores at most one block for each value written.
\\

We prove that in the absence of server to server communication and with up to $f$ Byzantine servers, if the writer can crash then implementing an auditable register that ensures completeness requires at least $4f+1$ servers. Our result is proved for a safe register as defined in \cite{L86} (which is weaker than an atomic one). This result does not depend on  the communication reliability.

We consider an auditable safe register, which is a safe register extended with the audit operation 
as defined in section \ref{property:completeness}.
A safe register ensures that if there is no write operation concurrent with a read operation $\op{}$, $\op{}$ returns the last value written in the register.

\begin{theorem}
\label{theorem:async:imopssibilityAudit}
No algorithm $\mathcal{P}$ implements an auditable safe register in an asynchronous system with $n<4f+1$ servers if the writer can crash and there is no server to server communication. 
\end{theorem}\spaceLemmaProof

\begin{proofT}
Let us proceed by contradiction, assuming that $\mathcal{P}$ exists. In particular, we consider the case of $n=4f$.

Consider an execution where the writer $\pw$ completes a write operation $\op{w}$, and after $\op{w}$ returns, a correct reader $\pr$ invokes a read operation $\op{r}$ which completes.
Let $v_1$ be the value written by $\op{w}$, since $\mathcal{P}$ exists, then $\op{r}$ returns $v_1$. Otherwise, we violate the safety property of the register.

As $\mathcal{P}$ ensures the liveness property and that there are $f$ Byzantine processes, we have that $\pw$ cannot wait for more than $n-f=3f$ acknowledgments from servers before completing $\op{w}$, i.e., $\pw$ cannot wait for more than $2f$ acknowledgments from correct servers before terminating.

Let us separate servers in three groups, $S_1$, $S_2$ and $S_3$ with $|S_1|=2f$, $|S_2|=f$ and $|S_3|=f$. Servers in $S_1$ and $S_2$ are correct, while servers in $S_3$ are Byzantine.

Let $\pw$ crash after $\op{w}$ terminates but before any servers in $S_2$ receive their block for $v_1$.
Since servers do not communicate with each other and that $\pw$ crashed, we can consider that no server in $S_2$ ever receives the blocks for $v_1$. 
Then only $2f$ correct servers, the ones in $S_1$ have a block for $v_1$. 
Since we cannot rely on Byzantine servers and each server stores at most one block, $\pr$ can collect at most $2f$ blocks for $v_1$. According to our hypothesis, $\mathcal{P}$ respect the safe semantic. Thus, $\pr$ is able to read the value by collecting only $2f$ different blocks.
However, according to Theorem \ref{theorem:impossible:bessaniCompleteness}, doing so $\mathcal{P}$ does not provide completeness, which is a contradiction.

\renewcommand{\toto}{theorem:async:imopssibilityAudit}

\end{proofT}


%% file: asyncUnboundedSolution.tex

We provide an
algorithm that implements a Single-Writer/Multi-Reader strong auditable wait-free atomic register. 

According to the impossibility result given by Theorem \ref{theorem:impossible:bessaniCompleteness}, the writer uses information dispersal techniques, with $\tau=2f+1$.
Our solution requires $3f+1$ servers, which is optimal resilient \cite{MAD02}. 
According to the impossibility result given by Theorem \ref{theorem:async:imopssibilityAudit}, we consider server to server communication, more in particular, we consider that the writer communicates with servers using a reliable broadcast abstraction. 
However, this nullifies the effect of using information dispersal techniques to prevent Byzantine servers from accessing the value.
Indeed, all the servers would deliver the $n$ blocks and then could reconstruct the value. To address this issue, the writer encrypts each block with the public key of the corresponding server, such that only the $i-th$ server can decrypt the $i-th$ block with its private key.

\subsection{Description of the algorithm}

Messages have the following syntax: 
   $\langle TAG,payload\rangle$. $TAG$ represent the type of messages and $payload$ is the content of the messages.

\noindent{\bf Variables at writer side:}\\
    \noindent{\bf -} $ts$ is an integer which represents the timestamp associate to the value being written (or lastly written) into the register.
    
    \noindent{\bf -} $b_1, \dots, b_n$ are the blocks related to the value being written (or lastly written) into the register. It is such that the block in $b_i$ is encrypted with the public key of the server $\ps{i}$.

\noindent{\bf Variables at reader side:}\\
All the following variables (except $\nSeq$) are reset at each new read operation invocation.\\
    \noindent{\bf -}  $\nSeq $ is an integer which represents the sequence number of the read operation of the reader $\pr$. This value is incremented at each read invocation.
    
    {\color{black}\noindent{\bf -}  $\collectedValues$ is an array of $n$ sets of tuple (block, timestamp). The $i-th$ position stores all the blocks associated with their timestamps, received from server $\ps{i}$ in response to $\getval$ messages (if any).}
    
    \noindent{\bf -}  $\collectedTs$ is an array of $n$ lists of integers. In each position $i$, it stores the list of all the timestamp received from server $\ps{i}$ in response to $\getts$ messages (if any). 
    
    \noindent{\bf -}  $\smallestTimestamp$ is an integer of the smallest timestamp stored in $\collectedTs$ that is greater than $2f+1$ timestamps in $\collectedTs$.

\noindent{\bf Variables at audit side:}\\
     \noindent{\bf -} $\collectedLog$ is an $n$ dimension array that stores in each position $i$ the log received from server $\ps{i}$ in response to $\AUDIT$ messages (if any). This variable is reinitialized at each audit invocation.
     
     \noindent{\bf -} $\mathcal{E}_{\pr,ts}$ is a list that stores the proof attesting that the reader $\pr$ have read the value associated with timestamp $ts$.
     
     \noindent{\bf -} $E_A$ Is a list that stores all the tuples process-timestamp, of all the read operation detected by the audit operation. This variable is reinitialized at each audit invocation.

\noindent{\bf Variables at server side $\ps{i}$:}\\
     \noindent{\bf -} $\Rts$ is an integer, which is the current timestamp at $\ps{i}$. This value is used to prevent the reader to read an old value.
     
     \noindent{\bf -} $\RVal$ is a list of tuple (block, timestamp) storing all the block receive by server $\ps{i}$.
     
     \noindent{\bf -} $\Logi$  is a list of tuples reader ID, timestamp, signed either by the reader itself or by the writer. Those tuples are used as a proof that the reader effectively read.
     
     \noindent{\bf -} $\ReadsServer$  is a list of tuples, reader ID, $\nSeq$, that identifies all the pending read operations. 

\noindent{\bf Functions:} 
    \noindent{\bf -} {\sf \generatesBlock$(v)$.} This function, invoked by the writer $\pw$, takes as input $v$, the value to write. Using information dispersal techniques, it returns an array of $n$ encrypted blocks $[b_1,\ldots,b_n]$, one per server. The block $b_i$ is encrypted with the public key of the server $\ps{i}$. Furthermore, to reconstruct the couple value-timestamp from the blocks, any combination of at least $\tau=2f+1$  blocks are required, without what no information about $v$ can be retrieved.
    
    \noindent{\bf -}  {\sf \decrypt($b_{v_i}$).} This function, invoked by server $\ps{i}$, takes as input the $i-th$ encrypted block, and using the private key of $\ps{i}$, it decrypts the corresponding block and returns it.
    
    \noindent{\bf -}  {\sf \GetValue($\collectedValues)$} This function, invoked by the readers, takes as input $\collectedValues$, an array of $n$ lists of tuples blocks, timestamp. It returns a value if there are across the $n$ lists, $\tau$ different blocks corresponding to that value and bottom otherwise. If multiple values can be returned, then the function returns the one with the highest timestamp associated.   
    
\noindent{\bf The write operation (Fig \ref{alg:async:unbound:AtomicRegisterClientWrite} and Fig \ref{alg:async:unbound:AtomicRegisterServer}).}  At the beginning of the write operation, the writer increase its timestamp and generates $n$ blocks, one for each server, with the $i-th$ block encrypted with the public key of server $\ps{i}$. Then the writer broadcast a $\writets$ message to all servers that contains the timestamp and all the blocks. Once a correct server $\ps{i}$ receives a $\writets$ messages, $\ps{i}$ add  the timestamp and the decrypted $i-th$ block to $\RVal$. Then $\ps{i}$ updates its timestamp in $\Rts$. Finally, the server acknowledge the writer in an $\writetsack$. Once the writer receives $n-f$ ACKs from different servers, it terminates.

\noindent{\bf The read operation. (Fig \ref{alg:async:unbound:AtomicRegisterClientRead} and Fig \ref{alg:async:unbound:AtomicRegisterServer})} 
The read operation takes place in several phases.
Foremost, the reader starts by increasing its sequence number. This sequence number is included in each message the reader sends, such that only response from servers with the same sequence number are considered. 

In the first phase, the reader sends $\getts$ to message to all servers. When a correct server $\ps{i}$ receives such message, it sends its timestamp $\Rts$ to the reader. After collecting $n-f$ timestamp from different servers, the reader begins the next phase.

In the second phase, the reader sends the $\getval$ message that contains $\smallestTimestamp$, to all servers. Timestamp $\smallestTimestamp$ is a timestamp received in a $\TS$ message, such that it is greater than or equal to $2f+1$ other timestamps received. If there are more than one that respect this condition, then the reader selects the smallest timestamp that satisfies such condition (cf. line \ref{line:async:unbound:notOld} figure \ref{alg:async:unbound:AtomicRegisterClientRead}). When a correct server $\ps{i}$ receives a $\GetValUnbound$ message, 
then if the timestamp in $\Rts$ is greater than or equal to $\smallestTimestamp$ receives in the $\getval$ message, then $\ps{i}$ sends back $\smallestTimestamp$ in a $\SendValUnbound$ message. If $\Rts$ is smaller than $\smallestTimestamp$, $\ps{i}$ waits to receive such timestamp. When the reader have collected $f+1$ $\smallestTimestamp$ it starts the next phase.

In the third phase the reader sends a $\GetBlocks$ message to all servers with the timestamp collected $f+1$ at the preceding phases. When a correct server $\ps{i}$ receive such messages, it sends back the corresponding block. If it has not yet such block, it waits to receive it. When the reader receive $n-f$ responses, it returns.

\noindent{\bf The audit operation ((Fig \ref{alg:async:unbound:AtomicRegisterClientWrite} and Fig \ref{alg:async:unbound:AtomicRegisterServer})).}
Such operation is similar as the one describe in \cite{CB20}. When a process $\pa$ performed an Audit operation, it sends an $\AuditReq$ messages to all server. When a server receives an $\AuditReq$ messages, it sends back to $\pa$ an $\AuditResp$ with its log. Then $\pa$ stores in $\mathcal{E}_{\pr}$ all occurrence of $\pr$ in the different logs it receives. If a process $\pr$ occurs more than $t$ times in $\mathcal{E}_{\pr}$, then it is added in the response $E_A$ of the audit operation.

\subsection{The algorithm}
\label{sec:async:unbound:AtomicRegister}


\begin{algorithm}[tb]
    \input{algorithms/asynchronousCase/auditClientWriteAsync}
    \caption{Pseudo-code of the write and audit operations at writer $\pw$.}
    \label{alg:async:unbound:AtomicRegisterClientWrite}
\end{algorithm}
    
\begin{algorithm}[tb]
    \input{algorithms/asynchronousCase/auditClientReadAsync}
    \caption{Pseudo-code of the read operation at reader $\pr$.}
    \label{alg:async:unbound:AtomicRegisterClientRead}
\end{algorithm}

\begin{algorithm}[tb]
    \input{algorithms/asynchronousCase/auditServerAsync}

    \caption{Pseudo-code at server $\ps{i}$.}
    \label{alg:async:unbound:AtomicRegisterServer}
\end{algorithm}


\subsection{Proof}

In the following, we prove that Algorithms in section \ref{sec:async:unbound:AtomicRegister} solves the Auditable Atomic Register problem. We first show that it satisfies the atomicity property, then that it satisfies wait-freedom and finally the completeness and accuracy property.

In this section we use the following notation. Given an operation $\op{x}, x \in \{r, w, a\}$, we denote $\tbegin{x}$ the operation invocation time instant and as $\tend{x}$ as the response time. For each couple of operations $\op{x}$ and $\op{y}$ ($x\neq y$), we say that $\op{x}$ succeeds $\op{y}$ if and only if $\tbegin{\op{x}} > \tend{\op{y}}$. For conciseness, when $x$ is a write or read operation, we refer to the timestamp associated to a value written by a write operation or returned by a read operation as $\opts{x}$. Let us recall that, $\op{r}.ts$ it is not related to $\nSeq$, which is the sequence number associated to that read operation.

\subsubsection{Atomicity Proof}

According to line \ref{line:async:unbound:updateRts} figure \ref{alg:async:unbound:AtomicRegisterServer}
the following observation holds :

\begin{observation}
\label{observation:unbound:tsMonotonicallyIncrease}
A correct server $\ps{i}$ updates $\Rts$ with a new timestamp, only if this timestamp is greater than the previous one stored in $\Rts$. Hence, timestamps stored at correct servers increase monotonically
\end{observation}

\noindent{}
\begin{lemma}
\label{lemma:async:unbound:writerEndWriteServerTimestamp}
Let $\op{w}$ be a write operation with timestamp $ts$. After $\op{w}$ returns the value of $\Rts$ is greater than or equal to $ts$ in at least $f+1$ correct servers. 

\end{lemma}\spaceLemmaProof 

\begin{proofL}
$\op{w}$ terminates when the condition at line \ref{line:async:unbound:writeTsWaitAck} of figure \ref{alg:async:unbound:AtomicRegisterClientWrite} is evaluated to true. 
Hence, $\op{w}$ terminates only if the writer received  $(\writetsack,ts)$, from at least $2f+1$ different servers. 
A correct server, sends $(\writetsack,ts)$ (line \ref{line:async:unbound:writeAck} of figure \ref{alg:async:unbound:AtomicRegisterServer}) to the writer, if it receives $(\writets,ts,-,-,-)$ from the writer and after the execution of line \ref{line:async:unbound:updateRts} in Figure \ref{alg:async:unbound:AtomicRegisterServer}
Since there are at most $f$ of the $3f+1$ servers are Byzantine, once the write operation completes, the writer has received at least $f+1$  $(\writetsack,ts)$ from correct servers.
Observation \ref{observation:unbound:tsMonotonicallyIncrease} concludes the proof. 

\renewcommand{\toto}{lemma:async:unbound:writerEndWriteServerTimestamp}
\end{proofL}

\begin{lemma}
\label{lemma:async:unbound:readerEndWriteServerTimestamp}
Let $\op{r}$ be a complete read operation and let $ts$ the timestamp corresponding to the value it returns. At any time after $\op{r}$ returns the value of $\Rts$ is greater than or equal to $ts$ in at least $f+1$ correct servers. 
\end{lemma}\spaceLemmaProof

\begin{proofL}
Since $\op{r}$ terminates, it has satisfied the condition $\ValidBlock(ts)$ (line \ref{line:async:unbound:conditionread} figure \ref{alg:async:unbound:AtomicRegisterClientRead}). For this condition to be true, the reader must have received $2f+1$ different valid blocks for timestamp $ts$, piggybacked by $\VALS$ messages (lines \ref{line:async:unbound:valRepColectTs} figure \ref{alg:async:unbound:AtomicRegisterClientRead}) sent by different servers. A correct server sends such messages, only after it has $\Rts\geq ts$ (lines \ref{line:async:unbound:waitTsSendVal} and \ref{line:async:unbound:updateRts}).
Since there are at most $f$ Byzantine servers, and by observation \ref{observation:unbound:tsMonotonicallyIncrease}, the claims follow.

\renewcommand{\toto}{lemma:async:unbound:readerEndWriteServerTimestamp}
\end{proofL}

\begin{lemma}
\label{lemma:async:unbound:W-Ratomicity}
Let $\op{w}$ be a complete write operation with timestamp $ts$ and let $\op{r}$ be a complete read operation by a correct process $\pr$ associated to a timestamp $ts'$. If $\op{r}$ succeeds $\op{w}$ in real-time order then $ts' \geq ts$
\end{lemma}\spaceLemmaProof 

\begin{proofL}
Since $\op{r}$ returns a value associated to timestamp $ts'$, the condition $\notOld(ts')$ (line \ref{line:async:unbound:conditionread} figure \ref{alg:async:unbound:AtomicRegisterClientRead}) is satisfied. 
In the following, we show that $\notOld(ts')$ true implies that $ts'\geq ts$.

Let us consider that $\notOld(ts')$ is true. 
As $\collectedTs$ is reinitialized at the beginning of each new read operation, $\notOld(ts')$ is true, if all timestamps receives from at least $2f+1$ different servers piggybacked by $\SendValUnbound$ (line \ref{line:async:unbound:valRepColectTs} figure \ref{alg:async:unbound:AtomicRegisterClientRead}) or $\TS$ (line \ref{line:async:unbound:valRepColectTs} figure \ref{alg:async:unbound:AtomicRegisterClientRead}) messages are smaller than or equal to $ts'$.
According to Lemma \ref{lemma:async:unbound:writerEndWriteServerTimestamp}, as $\op{w}$ terminates, the content of $\Rts$ is greater than or to equal  $ts$ in at least $f+1$ correct servers.
So, during $\op{r}$, in response to $(\getval,\nSeq)$ (line \ref{line:async:unbound:sendVal} figure \ref{alg:async:unbound:AtomicRegisterServer}), the reader can collect at most $2f=n-(n-2f)$ messages for a timestamp smaller than $ts$. Thus, $\notOld()$ always remain false for any timestamp smaller than $ts$, and since $\notOld(ts')$ is true,  $ts'\geq ts$.

\renewcommand{\toto}{lemma:async:unbound:W-Ratomicity}

\end{proofL}

\begin{lemma}
\label{lemma:async:unbound:returnWrittenValue}
If a complete read operation $\op{r}$, invoked by a correct process $\pr$,  returns a value corresponding to a timestamp $ts>0$, then it exists a write operation $\op{w}$, with timestamp $ts$, that starts before $\op{r}$ terminates.

\end{lemma}\spaceLemmaProof 
\begin{proofL}
Since $\op{r}$ terminates, it has satisfied the condition $\ValidBlock(ts)$ (line \ref{line:async:unbound:conditionread} of Figure \ref{alg:async:unbound:AtomicRegisterClientRead}). Thus, the reader received $2f+1$ distinct valid blocks for timestamp $ts$, piggybacked by $\SendBlock$  messages (lines \ref{line:async:unbound:receiveGetBlocks} and of Figure \ref{alg:async:unbound:AtomicRegisterClientRead}) sent by distinct servers. As all the variables are reinitialized at the beginning of a read operation, and as there are at most $f$ Byzantine servers, at least $f+1$  correct servers sent a block to $\pr$ during the execution of $\op{r}$. 
A correct server $\pss$ sends a block corresponding to a timestamp $ts>0$ only after it has received the corresponding $\writets$ message from the writer; thus, after the invocation of a write operation $\op{w}$ for timestamp $ts$. It follows that $\op{w}$ began before $\op{r}$ completes.

\renewcommand{\toto}{lemma:async:returnWrittenValue}
\end{proofL}

\begin{lemma} 
\label{lemma:async:unbound:R-Ratomicity}
Let $\op{r}$ be a complete read operation invoked by a correct process $\pr$, and let $\op{r'}$ be a complete read operation invoked by a correct process $\pr'$ ($\pr$ and $\pr'$ may be the same process). Let $ts$ and $ts'$ be the timestamps associated with $\op{r}$ and $\op{r'}$ respectively. If $\op{r'}$ succeeds $\op{r}$  in real-time order then $ts' \geq ts$.

\end{lemma}\spaceLemmaProof 

\begin{proofL}
The proof follows the same approach as the proof of Lemma \ref{lemma:async:unbound:W-Ratomicity}.

Let $ts'$ be $\op{r'}.ts$ and let $ts$ be $\op{r}.ts$.
As $\op{r'}$ returns a value associated to a timestamp $ts'$, the conditions $\ValidBlock(ts')$ and $\notOld(ts')$ are true for timestamp $ts$ (line \ref{line:async:unbound:conditionread} figure \ref{alg:async:unbound:AtomicRegisterClientRead}). 
In the following, we show that if condition $\notOld(ts')$ implies that $ts'\geq ts$.

Let us consider that $\notOld(ts')$ is true. 
As $\collectedTs$ is reinitialized at the beginning of each new read operation, $\notOld(ts')$ is true, if all timestamps receives from at least $2f+1$ different servers piggybacked by $\SendValUnbound$ (line \ref{line:async:unbound:valRepColectTs} figure \ref{alg:async:unbound:AtomicRegisterClientRead}) or $\TS$ (line \ref{line:async:unbound:valRepColectTs} figure \ref{alg:async:unbound:AtomicRegisterClientRead}) messages are smaller than or equal to $ts'$.
According to Lemma \ref{lemma:async:unbound:readerEndWriteServerTimestamp}, as $\op{r}$ returns for timestamp $ts$, then at server side, the content of $\Rts$ is greater than or to equal  $ts$ in at least $f+1$ correct servers.
So the reader can collect at most $2f=n-(n-2f)$  timestamps smaller than $ts$. Thus, $\notOld$ always remain false for any timestamp smaller than $ts$, hence $ts'\geq ts$.

\renewcommand{\toto}{lemma:async:unbound:R-Ratomicity}
\end{proofL}

\begin{lemma} 
\label{lemma:async:unbound:R-Watomicity}
Let $\op{r}$ be a complete read operation with timestamp $ts$ and let $\op{w}$ be a complete write operation by $\pw$ associated to a timestamp $ts'$. 
If $\op{w}$ succeeds $\op{r}$ in real-time order then $ts' \geq ts$
\end{lemma}\spaceLemmaProof 

\begin{proofL}
We proceed considering first the case in which $ts>0$ and then the case in which $ts=0$.
From Lemma \ref{lemma:async:unbound:returnWrittenValue}, if $ts>0$ then it exists $\op{w'}$ with timestamp $ts$ that starts before the end of $\op{r}$. Considering that, there is a unique writer and that its execution is sequential, then $\op{w'}$ terminates before $\op{w}$ starts. As timestamps growth monotonically (Observation \ref{observation:unbound:tsMonotonicallyIncrease}), 
$ts'>ts$.

Consider now the case $ts=0$. As timestamps grow monotonically (Observation \ref{observation:unbound:tsMonotonicallyIncrease}), and the initial value of the timestamp is $0$ (line \ref{line:async:unbound:initTs} of Figure \ref{alg:async:unbound:AtomicRegisterClientWrite}) then all write operations have their timestamp greater than $0$. In particular this is true for $\op{w}$, such that $ts'>ts$, which concludes the proof.
\renewcommand{\toto}{lemma:async:unbound:R-Watomicity}
\end{proofL}

Let $E$ be any execution of our algorithm and let $H$ be the corresponding history. 
We construct ${\sf complete}(H)$ by removing all the invocations of the read operations that have no matching response and by completing a pending $write(v)$ operation if there is a complete read operation that returns $v$. Observe that only the last write operation of the writer can be pending.

Then, we explicitly construct a sequential history $\pi$ containing all the operations in ${\sf complete}(H)$. 
First we put in $\pi$ all the write operations according to the order in which they occur in $H$, because write operations are executed sequentially by the unique writer, this sequence is well-defined. Also this order is consistent with that of the timestamps associated with the values written.

Next, we add the read operations one by one, in the order of their response in $H$. A read operation that returns a value with timestamp $ts$ is placed immediately before the write that follows in $\pi$ the write operation associated to $ts$ (or at the end if this write does not exist). By construction of $\pi$ every read operation returns the value of the last preceding write in $\pi$. 
It remains to prove that $\pi$ preservers the real-time order of non-overlapping operations.
\\
\begin{theorem} \label{theorem:async:unbound:linearizable}
Let $\op{1}$ and $\op{2}$ be two operations in $H$. If $\op{1}$ ends before the invocation of $\op{2}$ then $\op{1}$ precedes $\op{2}$ in $\pi$.

\end{theorem}\spaceLemmaProof

\begin{proofT}

Since $\pi$ is consistent with the order of timestamps, we have to show that the claim is true for all operations with different timestamps.

There are four possible scenarios: 
$\op{1}$ and $\op{2}$ are respectively a write and a read operation, then the claim holds by Lemma \ref{lemma:async:unbound:W-Ratomicity}.
$\op{1}$ and $\op{2}$ are two reads operations, then the claim holds by Lemma \ref{lemma:async:unbound:R-Ratomicity}. 
$\op{1}$ and $\op{2}$ are respectively a read and a write operation, then the claim holds by Lemma \ref{lemma:async:unbound:R-Watomicity}.
If $\op{1}$ and $\op{2}$ are two write operations the claim holds by the Observation \ref{observation:unbound:tsMonotonicallyIncrease}.

\renewcommand{\toto}{theorem:async:linearizable}
\end{proofT}

\subsubsection{Liveness Proof}

\begin{lemma}
\label{lemma:async:unbound:writeliveness}
If $\pw$ is correct (if the writer don't crash), then each write operation invoked by $\pw$ eventually terminates.
\end{lemma}\spaceLemmaProof

\begin{proofL}
The write operation has the following structure. The writer broadcasts a  $\writets$ message to all servers (line \ref{line:async:unbound:writeTs} of Figure \ref{alg:async:unbound:AtomicRegisterClientWrite}) and waits $n-f$ ACKs (line \ref{line:async:unbound:writeTsWaitAck} of Figure \ref{alg:async:unbound:AtomicRegisterClientWrite}) from different servers before terminate.

Since $\pw$ is correct, the channel communications properties assure that all correct servers deliver the message $\writets$ broadcast by $\pw$. Considering that: (i) servers do not apply any condition to send back $\writetsack$ messages (line \ref{line:async:unbound:writeAck} of Figure \ref{alg:async:unbound:AtomicRegisterServer}), and that (ii) at most $f$ servers can be faulty, $\pw$ always receives the $n-f$ $\writetsack$ replies from correct servers necessary to stop waiting, which concludes the proof.

\renewcommand{\toto}{lemma:async:unbound:writeliveness}
\end{proofL}

\begin{lemma}
\label{lemma:async:unbound:WriteRtsUpdate}
Let $\op{w}$ be a complete write operation that writes $v$ with timestamp $ts$. If a correct server $\pss$ updates $\Rts$ with $ts$, then all correct servers eventually adds the block corresponding to $v$ in  $\RVal$.

\end{lemma}\spaceLemmaProof  

\begin{proofL}
Let us recall that a correct server updates $\Rts$ with $ts$ only upon receiving a $\writets$ message (line \ref{line:async:unbound:receiveWriteVal} figure \ref{alg:async:unbound:AtomicRegisterServer}).
When a server updates $\Rts$ with $ts$, it also add to $\RVal$ (line \ref{line:async:unbound:updateRValWriteVal} figure \ref{alg:async:unbound:AtomicRegisterServer}) the associate block it receives in the $\writets$ message from the writer (line \ref{line:async:unbound:receiveWriteVal} figure \ref{alg:async:unbound:AtomicRegisterServer}). As there is a reliable broadcast between the writer and servers, eventually all correct servers receive the $\writets$ message and updates $\RVal$ with their block associate to $ts$.

\renewcommand{\toto}{lemma:async:unbound:WriteRtsUpdate}
\end{proofL}

\begin{lemma}
\label{lemma:async:unbound:smallestTimestampRts}
When a correct reader $\pr$ receives the response to $\getts$ from all correct servers, in at least one correct server $\Rts\geq\smallestTimestamp$.
\end{lemma}\spaceLemmaProof 

\begin{proofL}
By contradiction, assume that in all correct servers $\Rts<\smallestTimestamp$. 

Then, in response to $\getts$ messages, all correct servers send their $\Rts$, all inferior to $\smallestTimestamp$. We note $tsMax$ the greatest timestamp receives by $\pr$ from correct servers. Then, it exists in $\collectedTs$ $2f+1$ timestamp $\leq tsMax$. By assumption, $tsMax<\smallestTimestamp$, which is in contradiction with the condition line \ref{line:async:unbound:notOld}.

\renewcommand{\toto}{lemma:async:unbound:smallestTimestampRts}
\end{proofL}

\begin{lemma}
\label{lemma:async:unbound:ReadTermination}
A read operation $\op{r}$ invoked by a correct process $\pr$ always terminates.
\end{lemma}\spaceLemmaProof  

\begin{proofL}
First, observe that if $\pr$ satisfies the conditions at line \ref{line:async:unbound:conditionread} figure \ref{alg:async:unbound:AtomicRegisterClientRead}, $\pr$ terminates.
Then, let us show by construction that those conditions are necessarily satisfied. 

A correct reader $\pr$ starts the read operation, after it reinitialized all the variables. Then, $\pr$ sends a $\getts$ messages to all the servers. Consider the moment $\pr$ receives the response from the $2f+1$ correct servers, and sends a $\GetValUnbound$ message to all the servers for timestamp $\smallestTimestamp$. Notice that according to Lemma \ref{lemma:async:unbound:smallestTimestampRts} at least one correct server has set $\Rts$ to $\smallestTimestamp$. As at least one correct server set $\Rts$ to $\smallestTimestamp$, then from the reliable broadcast, eventually all servers will set $\Rts$ to $\smallestTimestamp$. Then, the reader will eventually receive the $2f+1$ response from correct servers such that the condition line \ref{line:async:unbound:conditionGetBlock} is satisfied. 

Then the reader sends $\GetBlocks$ messages for timestamp $\smallestTimestamp$ to all servers. As at least one correct server set $\Rts$ to $\smallestTimestamp$, from Lemma \ref{lemma:async:unbound:WriteRtsUpdate}, eventually all correct servers have in $\RVal$ the block associate with timestamp $\smallestTimestamp$. Then, in response to $\GetBlocks$, all correct servers can send their block corresponding to timestamp $ts$ and the condition $\ValidBlock$ is satisfied, such that the reader can return for the value-timestamp pair corresponding to timestamp $ts$.

\renewcommand{\toto}{lemma:async:unbound:ReadTermination}
\end{proofL}

\subsubsection{Auditability}

\begin{lemma}
\label{lemma:async:unbound:Completeness}
Algorithm presented figure \ref{alg:async:unbound:AtomicRegisterClientWrite} to \ref{alg:async:unbound:AtomicRegisterServer} 
solves the completeness property
\end{lemma}\spaceLemmaProof 
\begin{proofL}
 For a reader $\pr$ to returns for a valid value $v$ with timestamp $ts$, then $\pr$ receives at least $\tau$ messages from different servers (line \ref{line:async:unbound:conditionread} figure \ref{alg:async:unbound:AtomicRegisterClientRead}) with the block corresponding to $v$.
 A correct server sends a block with associate timestamp $ts$ to a reader $\pr$ (line \ref{line:async:unbound:sendVal}, figure \ref{alg:async:unbound:AtomicRegisterServer}), only after it adds to its log the reader $\pr$ associate with timestamp $ts$, line \ref{line:async:unbound:logServRval} figure \ref{alg:async:unbound:AtomicRegisterServer}. Thus, if a correct server $\pss$ sends a block with associate timestamp $ts$ to a reader $\pr$, $\pss$ stores $\pr$ ID associate with $ts$ in its log. If $\tau\geq 2f+1$, since there is at most $f$ Byzantine, in the worst case, at least $\tau-f\geq f+1$ correct servers, denoted $P_C$, records $\op{r}$ in their logs. 

Let $\op{a}$ be an audit operation, invoked by a process $\pa$, that starts after $\pr$ returns.
So when $\op{a}$ starts, $\pr$ is in PC's log. Then, $\pa$ waits $2f+1$ responses (line \ref{line:async:unbound:waitAuditResponse} figure \ref{alg:async:unbound:AtomicRegisterClientWrite}) after sending $\AUDIT$ request (line \ref{line:async:unbound:auditRequest} figure \ref{alg:async:unbound:AtomicRegisterClientWrite}) to servers. As there is at most $f$ Byzantine servers, $\pa$ gets the responses from at least $f+1$ ($n-2f$) correct servers. In particular, $\pa$ get at least one response from a server in $P_C$. Finally, with $t\leq\tau-2f=1$, $\pr$ and is reported by $\op{a}$ to have read the value associate to timestamp $ts$.

\renewcommand{\toto}{lemma:async:unbound:Completeness}
\end{proofL}

\begin{lemma}
\label{lemma:async:unbound:CompletenessCollusion}
Algorithm presented figure \ref{alg:async:unbound:AtomicRegisterClientWrite} to \ref{alg:async:unbound:AtomicRegisterServer} 
solves the completeness (with collusion) property
\end{lemma}\spaceLemmaProof 
\begin{proofL}
 For a reader $\pr$ to returns for a valid value $v$ with timestamp $ts$, then $\pr$ receives at least $\tau$ messages from different servers (line \ref{line:async:unbound:conditionread} figure \ref{alg:async:unbound:AtomicRegisterClientRead}) with the block corresponding to $v$. Notice that if $\pr$ is faulty, it can also receive those blocks not directly from the servers but for some other faulty process in $\FaultySet$. However, those other faulty process at some point must have receives those blocks from the servers.
 
 A correct server sends a block with associate timestamp $ts$ to a reader $\pr'$ (line \ref{line:async:unbound:sendVal}, figure \ref{alg:async:unbound:AtomicRegisterServer}), only after it adds to its log the reader $\pr'$ associate with timestamp $ts$ (line \ref{line:async:unbound:logServRval} figure \ref{alg:async:unbound:AtomicRegisterServer}). Thus, if a correct server $\pss$ sends a block with associate timestamp $ts$ to a reader $\pr'$, $\pss$ stores $\pr'$ ID associate with $ts$ in its log. If $\tau\geq 2f+1$, since there is at most $f$ Byzantine, in the worst case, at least $\tau-f\geq f+1$ correct servers, denoted $P_C$, records in their logs directly the reader $\pr$, or if $\pr$ is faulty, some faulty process in $\FaultySet$.

Let $\op{a}$ be an audit operation, invoked by a process $\pa$, that starts after $\pr$ returns.
So when $\op{a}$ starts, $\pr$ is in PC's log. Then, $\pa$ waits $2f+1$ responses (line \ref{line:async:unbound:waitAuditResponse} figure \ref{alg:async:unbound:AtomicRegisterClientWrite}) after sending $\AUDIT$ request (line \ref{line:async:unbound:auditRequest} figure \ref{alg:async:unbound:AtomicRegisterClientWrite}) to servers. As there is at most $f$ Byzantine servers, $\pa$ gets the responses from at least $f+1$ ($n-2f$) correct servers. In particular, $\pa$ get at least one response from a server in $P_C$. Finally, with $t\leq\tau-2f=1$, if $\pr$ is correct, $\pr$ is reported by $\op{a}$ to have read the value associate to timestamp $ts$, otherwise some faulty process in $\FaultySet$ are.

\renewcommand{\toto}{lemma:async:unbound:CompletenessCollusion}
\end{proofL}

\begin{lemma}
\label{lemma:async:unbound:StrongAccuraySignedMessage}
Algorithm presented figure \ref{alg:async:unbound:AtomicRegisterClientWrite} to \ref{alg:async:unbound:AtomicRegisterServer} with $t \geq 1$ solves the strong accuracy property
\end{lemma}\spaceLemmaProof 

\begin{proofL}
We have to prove that a correct reader $\pr$ that never invoked a read operation cannot be reported by an audit operation. With $t=1$, a reader $\pr$ is reported by an audit operation if one server respond to the $\AUDIT$ message with a correct record of $\pr$ in its log. Thanks to the use of signature, a false record cannot be created by a Byzantine server. The signature used to attest the validity of a record are of two kind. If a correct server add $\pr$ in its log before responding to $\GetValUnbound$ messages (line \ref{line:async:unbound:logServRval}, figure \ref{alg:async:unbound:AtomicRegisterServer}), then it uses the reader signature. So for a process to have a valid record to $\pr$ in its log, the process $\pr$ must have sent $\GetValUnbound$ messages to some servers,  i.e. $\pr$ must have invoked a read operation.

\renewcommand{\toto}{lemma:async:unbound:StrongAccuraySignedMessage}

\end{proofL}

\begin{theorem}
\label{theorem:async:unbound:StrongauditabilitySignedMessage}
Algorithm presented figure \ref{alg:async:unbound:AtomicRegisterClientWrite} to \ref{alg:async:unbound:AtomicRegisterServer} with $n=3f+1$, $\tau = 2f+1$ and $t=1$ solves the strong auditability property
\end{theorem}

\begin{proofT}
Directly from Lemma \ref{lemma:async:unbound:StrongAccuraySignedMessage} and Lemma \ref{lemma:async:unbound:Completeness} with $t=1$ and $\tau=2f+1$.

\renewcommand{\toto}{theorem:async:unbound:StrongauditabilitySignedMessage}
\end{proofT}

%% file: algorithms/asynchronousCase/auditClientWriteAsync.tex
{
\begin{algorithmic}[1]

\Statex Initialization
\Statex $ ts \gets 0$ \label{line:async:unbound:initTs}
\Statex $ b_i \gets \perp\, \forall i  \in [1,n]$
\Statex $ \writerAcks[i] \gets \perp,\, \forall i  \in [1,n]$

\Statex

\START[Write]($v$)\{
\State{$ts \gets ts+1$} \label{line:async:unbound:incts}
\State{$[b_1,\ldots,b_n]\gets \generatesBlock(v) $} 

\START[\texttt{broadcast}]$(\writets,ts,[b_1,\ldots,b_n])$ to all servers
\label{line:async:unbound:writeTs}
\EndPhase
\State{\textbf{wait until} $|\{x:\writerAcks[x]= ts\}|\geq n-f$}
\label{line:async:unbound:writeTsWaitAck}

\State{\textbf{return}}
\EndPhase
\Statex \}
\Statex

\START[upon \texttt{receive}] ($\writetsack,t$) from server $\ps{i}$
\START[if] ($ts=t$) \,\{ $\writerAcks[i]\gets t$\}
\EndPhase
\EndPhase

\Statex

\START[Audit()]\{
\label{line:async:unbound:auditStart}
\Statex $\collectedLog[i] \gets \emptyset, \forall i \in [1,n] $
\START[for] $i \in  [1,n]$ \texttt{\textbf{send}}$(\AuditReq)$ to server $\ps{i}$
\label{line:async:unbound:auditRequest}
\EndPhase
\State \textbf{wait until} $ |\{i:\collectedLog[i]\not=\emptyset\}|\geq n-f$
\label{line:async:unbound:waitAuditResponse}
\START[for] \textbf{all}$(\pr,ts,\nSeq)\in \bigcup_{i\in[1,n]}\collectedLog[i]$ 
\START[for] $1\leq k\leq n$  \textbf{if} $(\pr,ts,\nSeq)_{\sigma_{\pr}} \in \collectedLog[k]$  $\land(\pr,ts,\nSeq)_{\sigma_{\pr}}\not \in E_A$
\State\textbf{then} $E_A\gets E_A\cup (\pr,ts,\nSeq)_{\sigma_{\pr}}$
\EndPhase
\label{line:async:unbound:conditionReportAudit}
\EndPhase
\State{\textbf{return} $E_A$} 
\label{line:async:unbound:auditEnd}

\EndPhase
\Statex \}

\START[upon \texttt{receive}] ($\AuditResp, \Logi$) from server $\ps{i}$
\State \textbf{if} $\Logi=\emptyset$ \textbf{then} \{ $\collectedLog[i]\gets \perp$ \}
\State \textbf{else} $\collectedLog[i]\gets \Logi$
\EndPhase

\end{algorithmic}
}

%% file: algorithms/asynchronousCase/auditClientReadAsync.tex
{
\begin{algorithmic}[1]
\Statex \textbf{Definitions:}
\Statex $\ValidBlock(ts) \triangleq  {\sf \GetValue} (\collectedValues)=(v,ts)$ 
\Statex \textcolor{black}{ $\notOld( ts) \triangleq |\{i:min(\collectedTs[i])\leq ts\}|\geq 2f+1$ }
\Statex

\Statex \textbf{Initialization :}
\Statex $\nSeq \gets 0$
\Statex $\collectedValues[i] \gets (\perp,\perp), \forall i \in [1,n] $
\Statex $\collectedTs[i] \gets \emptyset, \forall i \in [1,n] $
\Statex $\smallestTimestamp \gets \perp$
\Statex
\START[Read()]\{
\State{$\nSeq \gets \nSeq+1$}
\State $\collectedValues[i] \gets (\perp,\perp), \forall i \in [1,n] $
\State $\collectedTs[i] \gets \emptyset, \forall i \in [1,n] $
\State $\smallestTimestamp \gets \perp$

\START[for] $i \in  [1,n]$ \texttt{\textbf{send}}$(\getts,\nSeq)$ to server $\ps{i}$
\label{line:async:unbound:sendGetTs}

\EndPhase
\State \textbf{wait until}
$(|\{ts\in\collectedTs:ts=\smallestTimestamp\}|\geq f+1)$
\label{line:async:unbound:conditionGetBlock}

\START[for] $i \in  [1,n]$ \texttt{\textbf{send}}$(\GetBlocks,(\pr,\smallestTimestamp,\nSeq)_{\sigma_{\pr}})$  to server $\ps{i}$
\label{line:async:unbound:sendGetBlocks}
\EndPhase

\State \textbf{wait until}
$(\ValidBlock(ts)\land\notOld(ts))$
\label{line:async:unbound:conditionread}

\label{line:async:unbound:endReadMessage}
\State \textbf{return}  $\GetValue(\collectedValues)$
\label{line:async:unbound:readReturn}

\EndPhase
\Statex
\Statex

\START[upon \texttt{receive}]{ (\TS,ts,num) from server $\ps{i}$}
\START[if] ($num=\nSeq$)
    \State $\collectedTs[i]\gets \collectedTs[i]\cup ts$
    \label{line:async:unbound:tsRepColectTs}
    \START[if] $(|x:\collectedTs[x]\not=\perp\}|\geq n-f$
    \State\textcolor{black}{ $\smallestTimestamp \gets min(\{ts\in \collectedTs$ $|$ $\notOld(ts)\})$}
    \label{line:async:unbound:notOld}
    \START[for] $i \in  [1,n]$ \texttt{\textbf{send}} $(\GetValUnbound,\smallestTimestamp,\nSeq)$  to server $\ps{i}$
    \label{line:async:unbound:sendGetVal}
    \EndPhase
    \EndPhase
    \EndPhase
\EndPhase

\Statex

\START[upon \texttt{receive}] ($\SendValUnbound,ts,num$) from server $\ps{i}$
\START[if] ($num=\nSeq$)
\{$\collectedTs[i]\gets \collectedTs[i]\cup ts$\}
\label{line:async:unbound:valRepColectTs}
    \EndPhase
\EndPhase
\Statex

\START[upon \texttt{receive}]{ ($\SendBlock,b_{i},ts,num$) from server $\ps{i}$}
\START[if] ($num=\nSeq$)
    \State $\collectedValues[i]\gets (b_i,ts)$
    \label{line:async:unbound:ColectBlock}
    \EndPhase
\EndPhase

\end{algorithmic}
}

%% file: algorithms/asynchronousCase/auditServerAsync.tex
{
\begin{algorithmic}[1]
\Statex Initialization
\Statex $\Rts_{i} \gets  0$
\Statex $\RVal_{i} \gets \emptyset$
\Statex $\Logi \gets \emptyset$
\Statex $\READERS_{i} \gets \emptyset$

\label{line:async:unbound:init}

\Statex

\Statex // Write Protocol Messages
\Comment{Pseudo code for server i}

\START[upon \texttt{receive}] ($\writets,ts,[b_1,\ldots,b_n]$) from writer $\pw$
\label{line:async:unbound:receiveWriteVal}
\State $\RVal_{i} \gets \RVal_{i} \cup (b_{i},ts)$
\label{line:async:unbound:updateRValWriteVal}
\START[if] {$\Rts_{i}<ts$}
\{$\Rts_{i} \gets ts$\}
\label{line:async:unbound:updateRts}
\EndPhase
\State \texttt{\textbf{send}} $\writetsack$ to $\pw$
\label{line:async:unbound:writeAck}
\EndPhase

\Statex

\Statex // Read Protocol Messages
\START[upon \texttt{receive}] ($\getts,\nSeq$) from reader $\pr$
\State{\texttt{\textbf{send}}($\TS,\Rts,\nSeq)$)}
\label{line:async:unbound:sendTs}
\EndPhase

\Statex 

\START[upon \texttt{receive}] ($\GetValUnbound,ts,\nSeq $) from reader $\pr$
\START[wait for] ($\Rts \geq ts )$
\label{line:async:unbound:waitTsSendTs}
\State \texttt{\textbf{send}}($\SendValUnbound,ts,\nSeq$) to $\pr$
\label{line:async:unbound:sendTs2}
\EndPhase
\EndPhase

\Statex

\START[upon \texttt{receive}] $(\GetBlocks,(\pr,ts,\nSeq)_{\sigma(\pr)})$ from reader $\pr$
\label{line:async:unbound:receiveGetBlocks}
\START[wait for] $\Rts\geq ts$
\label{line:async:unbound:waitTsSendVal}
\State $\Logi \gets \Logi \cup
(\pr,ts,\nSeq)_{\sigma(\pr)}$
\label{line:async:unbound:logServRval}
\State \texttt{\textbf{send}}($\SendBlock,\RVal[ts],\nSeq$) to $\pr$
\label{line:async:unbound:sendVal}
\EndPhase
\EndPhase

\Statex

\Statex // Audit protocol Messages
\START[upon \texttt{receive}] ($\AuditReq$) from owner $\pR$ of R 
\State \texttt{\textbf{send}}($\AuditResp$,$\Logi$) to $\pR$
\label{line:async:unbound:senLog}

\EndPhase
\Statex

\end{algorithmic}{}
}

%% file: writer_quorum.tex
 In this section we explore the guarantees obtained by switching from a single writer than can crash to multiples writers, with some byzantines.\\
 We consider $N_w \geq 2 * f_w + 1$ writers are trying to write au unique value $v$ in an auditable distributed register.
 We present a protocol for the writers that is compatible with any write operation that completes in a single round like the one presented in \ref{sec:asynchAlgorithm}. 
\begin{enumerate}[label=\arabic*]
    \item Writers perform the secret sharing in a deterministic way\footnote{This can be achieved by using $v$ to seed a PRNG that the writers then use to perform the secret sharing}.
    \item Writers send to each server its share.
    \item Servers wait until they receive the same share from $f_w + 1$ distinct writers before accepting it.   
\end{enumerate}
Using this setup, the following properties arise :\\
\vspace{0.5cm}
\begin{theorem}
	Correct servers only accept valid shares.
\end{theorem}
\begin{proofT}
	A correct server waits until collecting $f_w + 1$ copy of the same share before committing it to its storage. Because at most $f_w$ writers can be byzantines, at least one correct writer communicated the share to our server and as such, the share is necessarily correct.  
\end{proofT}
\vspace{0.5cm}
\begin{theorem}
    If one correct server accepts a share then, \\emph{eventually}, every correct server will accept their share.
\end{theorem}
\begin{proofT}
    From the previous theorem we know that the accepted share is valid. From this we know that the writers are in the process of sending shares to every server. Because there are more than $f_w + 1$ correct writers and we are operating in an eventually consistent network, eventually, every server will receive $f_w + 1$ times their share and thus every correct server will \emph{eventually} accept their share.
\end{proofT}
\vspace{0.5cm}
With this, we basically reduced the writing process to one with a single correct writer. We therefore removed the need for protocols to be crash-tolerant, while keeping every other property they might have. In particular, this means that reader of distributed registers implementing the above write sequence know that values that they read from the register must originate from correct writers by construction, hence the name \textbf{Valid Reads}.